# Molecular Dynamics Study of sp-Defect Migration in Odd Fullerene: Possible Role in Synthesis of Abundant Isomers of Fullerenes

Alexander S. Sinitsa[a], Irina V. Lebedeva[b], Yulia G. Polynskaya[d], Andrey M. Popov[c*], and Andrey A. Knizhnik[d]

[a]National Research Centre "Kurchatov Institute", Kurchatov Square 1, Moscow 123182, Russia.

[b]CIC nanoGUNE BRTA, Avenida de Tolosa 76, San Sebastian 20018, Spain.

[c]Institute for Spectroscopy of Russian Academy of Sciences, Fizicheskaya Street 5, Troitsk, Moscow 108840, Russia.

[d]Kintech Lab Ltd., 3rd Khoroshevskaya Street 12, Moscow 123298, Russia.

**ABSTRACT:** To explain recent experiment showing the role of odd fullerenes in formation of abundant fullerene isomers a reactive molecular dynamics (MD) study has been performed. Three types of bond rearrangement reactions are found by MD simulations at 3000 K in odd fullerenes which contain an extra sp atom among all other $sp^2$ atoms. The first type is stochastic sp-defect migration analogous to exchange mechanism of adatom migration on a surface. The second type corresponds to changes in the ring configuration of the $sp^2$-structure assisted by the sp atom which can lead to annealing of seven-membered rings or separation of five-membered rings. The third type is formation of short-living one-coordinated atoms or two additional sp atoms. Annihilation of a pair of sp defects has been also observed in the MD simulations. It is shown that the frequency of sp-defect migration at a lower temperature, as estimated from performed density functional theory calculations of the barriers of sp-defect migration events, is sufficient to deliver the sp atom to defects of $sp^2$ structure during the fullerene formation time. Based on these results, we propose to supplement the self-organization paradigm of fullerene formation by the following four-stage atomistic mechanism of formation of abundant isomers of fullerenes: 1) attachment of single carbon atoms, 2) sp-defect migration to $sp^2$-structure defects, 3) $sp^2$-defect annealing assisted by the sp atom and 4) subsequent annihilation of pairs of sp defects.

## I. INTRODUCTION

It is commonly accepted that fullerene formation occurs via self-organization of an initially chaotic carbon system without precursors of certain structure. The detailed arguments in favour of this route are given in reviews[1,2]. However, after more than thirty years since the discovery of fullerenes[3], the puzzle of the high yield of certain abundant isomers of fullerenes (like $C_{60}$-$I_h$ fullerene with icosahedral symmetry and $C_{70}$-$D_{5h}$ fullerene) is not yet solved. Recent experiment[4] demonstrates the role of odd fullerenes in formation of abundant isomers of fullerenes. However, modern experimental techniques do not allow to obtain the information about structural changes during the fullerene synthesis with a high yield of abundant isomers. Here we propose to supplement the self-organization paradigm of fullerene formation by the four-stage atomistic mechanism of formation of abundant fullerene isomers including odd fullerenes and based on molecular dynamics (MD) simulations performed.

The primary argument in favour of the self-organization paradigm is that fullerenes can be produced not only by the traditional synthesis method using arc discharge between graphite electrodes[5] but also by different methods from various initial systems consisting of pure carbon material: laser ablation of higher carbon oxides produces rings $C_{18}$, $C_{24}$, and $C_{30}$, which merge into large clusters transformed afterwards into fullerenes[6,7]; transformation of bi- and tricyclic clusters to fullerenes has been revealed in the drift tube[8-11]; fullerenes containing hundreds of atoms are formed upon merging several $C_{60}$ fullerenes during ablation of the pure fullerene $C_{60}$ film[12]; transformation of a graphene flake to a fullerene on a surface under electron irradiation have been observed by tunnelling electron microscopy (TEM)[13], simulations devoted to transformation of a graphene flake to a fullerene should be also mentioned[14-18]; transformation of amorphous carbon cluster to a fullerene has been predicted by MD simulations[19]. Fullerene formation starting from various initial carbon systems by self-organization is expected as a fullerene is the ground state for a carbon system containing from several tens to several hundreds atoms[20-22].

The structure of a perfect fullerene without defects consists only of 5-membered and 6-membered rings (below referred as 5- and 6-rings). Let us discuss the adequacy of using the term 'fullerene' in the cases of various defects of fullerene structure studied in the present paper. Here we consider fullerenes with only one or few local defects of atomic structure involving only one or few atoms. The structure of the objects considered is far from the structure of cage-like carbon clusters in the same way as the structure of crystals with a low density of defects is far from the amorphous one. Thus we refer to the objects studied as fullerenes.

According to the isolated pentagon rule, abundant isomers of fullerenes consist only of 5- and 6-rings with 5-rings separated by 6-rings[23]. Here 4-, 7-, and 8-rings as well as adjacent 5-ring are referred as $sp^2$ defect of a fullerene if these rings consist only of $sp^2$ atoms. Numerous MD simulations devoted to self-organization of chaotic carbon systems such as vapour[24–31] or amorphous cluster[19] showed that just after complete formation of the $sp^2$ structure, the fullerene shell contains numerous structural defects such as 7-, 8-rings and other rings in addition to 5- and 6-rings. Defective fullerenes were also revealed in MD simulations of transformations of short carbon nanotubes with open ends[32,33], graphene flakes[17,18] and small nanodiamond clusters[34]. To explain annealing of $sp^2$ defects and abundance of certain isomers of fullerenes, it is necessary to consider reactions of structure transformation after the initial fullerene shell formation. Originally only reactions which preserve the $sp^2$ structure of the even fullerene (that is the system which contains an even number of atoms and only $sp^2$ atoms are present before and after the reaction) and can in principle lead to $sp^2$-defect annealing and selection of abundant isomers have been considered. Two types of such reactions have been proposed: isomerization reactions corresponding to bond rearrangements keeping the constant number of atoms in the fullerene and insertion/emission of the $C_2$ molecule. The simplest reaction of a bond rearrangement in fullerenes is the Stone-Wales (SW) reaction of bond rotation by about 90° with respect to the midpoint of the bond[35]. All the listed reactions, the SW reaction[36–38], $C_2$ molecule insertion[37–40], $C_2$ molecule emission[41], and insertion of $C_2$ fragment from $C_3$ molecule[37] have been considered as a part of the main atomistic mechanism of selection of abundant fullerene isomers. However, according to density functional theory (DFT) calculations, these reactions should have considerable activation barriers: the barrier is 5–7 eV for the SW reaction[38–46], barriers of insertion of the $C_2$ molecule into the $sp^2$ structure of the fullerene range from 1–2 eV[38,46] to 4–5 eV[47] and the $C_2$ emission needs in total about 12 eV[47]. In addition to consideration of energetics of single reactions, statistical results for $sp^2$-defect annealing in fullerenes via the SW-reaction[36] and $C_2$ molecule insertion and emission[40] have been obtained using the kinetic approach[36] and MD simulations[40].

Pristine fullerenes contain only an even number of $sp^2$ atoms (referred here as even fullerenes). However, if one carbon atom is added to or removed from an even fullerene, the shell structure will contain only $sp^2$ atoms except one or few atoms which have not $sp^2$ hybridization and located within one structural defect. Such fullerene shell is referred here as an odd fullerene. To overcome the mentioned problem of high barriers for the reactions of bond rearrangements in even fullerenes, a set of atomistic mechanisms where the attachment of an extra sp atom to an even fullerene can lead to selection of abundant isomers have been considered[4,42,43,46,48]. The possibility of such atomistic mechanisms with participation of the sp atom is excellently confirmed by the recent experiment where the formation of the abundant isomer $C_{70}$-$D_{5h}$ has been observed upon laser ablation of the $C_{60}$-$I_h$ fullerene simultaneously with amorphous $^{13}C$[4]. The analysis of the isotope distribution in the mass-spectrum of the obtained $C_{70}$-$D_{5h}$ fullerene shows that insertion of atomic carbon (along with insertion of $C_2$ molecules) should considerably contribute to the observed closed-shell fullerene growth. The experimental studies of the $C_{60}$-$I_h$ fullerene with $C^+$ collisions show that the barrier of formation of the $C^+_{61}$ fullerene is less than 0.5 eV[49]. Such a barrier can be even less for fullerenes with $sp^2$ defects. While pristine fullerenes consist of an even number of atoms, the self-organization concept does not give any preference to formation of clusters with an even number of atoms before the completion of fullerene shell formation (this statement is confirmed here by MD simulations performed). Therefore, formation of short-living odd fullerenes can be expected for diverse conditions of fullerene synthesis.

The following theoretical results have been obtained up to now for bond rearrangements in odd fullerenes in relation with formation of abundant fullerene isomers. Firstly, it was proposed that the presence of an extra sp atom considerably reduces the barrier of the SW reaction in fullerenes down to 1.3 – 4.3 eV[4,42,43,46,48]. Such SW reactions in fullerenes assisted by an extra sp atom have been considered in two cases: 1) when the extra sp atom located outside the fullerene shell before and after the reaction participates only in a single SW reaction[48] and 2) when the extra sp atom is considered as an adatom sitting at the center of the bond of the even fullerene (two types of such reactions have been studied in literature[4,42,43,46]). Secondly, according to the DFT calculations, insertion of a carbon atom into the even fullerene shell with formation of an sp atom in the place of the former bond is barrierless[46]. Thirdly, the possibility of attachment of two carbon atoms at different places of an even fullerene shell resulting in formation of the pristine $sp^2$ structure via annihilation of two sp atoms has been also proposed[4]. The path for growth of the $C_{68}$ fullerene into the $C_{70}$ fullerene including two insertions of a single carbon atom and one SW-reaction assisted by the sp atom has been considered using the DFT calculations and it has been found that the barriers along this path are less than 2 eV[46].

Thus, up to now energetics and structural changes in odd fullerenes have been considered only for single reactions of bond rearrangements related with defect annealing[4,42,43,46,48]. Moreover, the whole atomistic mechanism of formation of abundant isomers of fullerenes has not been developed since the problems of 1) how the sp atom gets close to a defect which can be annealed with assistance of the sp atom and 2) how the sp atom is removed from an odd fullerene upon formation of the abundant isomer of the odd fullerene have so far received little attention. In the present paper, we perform reactive empirical MD simulations to generate odd fullerenes by heating amorphous carbon clusters and to study various bond rearrangement reactions in odd fullerenes. Various bond rearrangement reactions which always take place close to the sp atom are observed including reactions which deliver a sp atom to a sp$^2$ defect such as 7-ring or adjacent 5-rings and referred to here as sp-defect migration events. The MD simulations are supplemented here by DFT calculations of the barriers of sp-defect migration events. We also study the correlation between the odd fullerene energy and the number of 4-, 7-, and 8-rings to show the possibility of annealing of these rings in odd fullerenes. Three ways to obtain an even fullerene after defect annealing assisted by the sp atom have been proposed 1) sp atom attaches an even fullerene immediately close to a sp$^2$-defect and is emitted just after defect annealing[48], 2) another single carbon atom attaches nearby the first sp atom with subsequent formation of the sp$^2$ structure through a single bond rearrangement reaction[46] and 3) two sp atoms attach at different places of the even fullerene, migrate, meet and annihilate[4]. The latter way is observed here in the MD simulations.

The energetics and isomerization kinetics of odd fullerenes is important not only for fullerene formation but also for any other processes where odd fullerenes are formed from even fullerenes. Such processes include, for example structural transformations under electron irradiation in TEM[50] and under cosmic ray irradiation in the interstellar medium[51], with knock-on removal of atoms from the fullerene structure. Among the processes in TEM where structural transformations of fullerenes are important, we can mention formation of a trilobate structure from three coalescing La@C$_{82}$ endofullerenes[52], transformation of polyhedral graphitic multi-layer nanoparticles into quasi-spherical onions[53,54] and formation of double-walled nanotubes from single-walled nanotubes filled with C$_{60}$ fullerenes [50,55].

The paper is organized the following way. In Section 2 we describe the calculation methods. Section 3 presents the results of MD simulations of bond rearrangement reactions in odd fullerenes and DFT calculations of the barriers of sp-defect migration events. Section 4 is devoted to the discussion and conclusions.

## II. COMPUTATIONAL DETAILS

**II.A Details of molecular dynamics simulations.** MD simulations have been carried out using the latest version of REBO-1990EVC potential[56]. This modified version of the first-generation Brenner potential accurately describes elastic energies of fullerenes, energies of carbon chains, graphene edges and vacancy migration in graphene, which makes it an adequate choice for simulations of structural rearrangements of a fullerene shell. The in-house MD-kMC (Molecular Dynamics – kinetic Monte Carlo) code[57] is used for the simulations. The integration is performed with the velocity Verlet integration algorithm[58,59] with the time step of 0.6 fs. The temperature is kept fixed by the Berendsen thermostat[60] with the relaxation time of 0.03 ps. Bond breaking and formation is detected through the analysis of the topology of the bond network using the "shortest-path" algorithm[61] every 25 ps. Two atoms are considered as bonded if the distance between them lies within 1.8 Å. The barriers of the sp-defect migration events for the used empirical potential are calculated by the nudged elastic band method[62].

To study the possible role of the sp-defect migration during the fullerene formation, we consider odd fullerenes formed spontaneously from an initially chaotic carbon system. Recently we performed the MD simulations of the transformation of amorphous carbon clusters which contain an even number of atoms into even fullerenes[19]. Here we applied the same procedure to generate amorphous carbon clusters and to obtain odd fullerenes by heating the amorphous carbon clusters with an odd number of atoms. The previous version of the REBO-1990EVC potential[19] has been used.

**II.B Details of DFT calculations.** To obtain energy characteristics of sp-defect migration events in C$_{69}$ fullerene we have performed spin-polarized DFT calculations using two codes, VASP[63] and Priroda[64], with the same Perdew-Burke-Ernzerhof (PBE) exchange-correlation functional[65] but different electron basis sets. The VASP code has been used previously to fit the parameters of the REBO-1990EVC potential[56], whereas the Priroda code allows us to calculate barriers of reactions with a much smaller computational effort. Thus, the VASP code is used in the present paper only to calculate energy changes of migration events, while the Priroda code is applied both for barriers and energy changes. In the Priroda code, the calculations are performed using a Gaussian all-electron double-$\zeta$ basis set (cc-pVDZ)[66]. The reaction pathways are preliminarily analyzed by scanning the potential energy surfaces along the reaction coordinate. The exact transition state (TS) structures are determined following the Berny algorithm[67]. To confirm the TS structure, intrinsic reaction coordinate (IRC) calculations[68] are carried out following the single imaginary mode. In the VASP code, the interaction of valence electrons with atomic cores is described by the projector augmented-wave method[69]. The maximal kinetic energy of the plane-wave basis set is 600 eV. A second-order Methfessel-Paxton smearing[70] with a width of 0.1 eV is applied. The calculations are performed at the Gamma point for the 20 Å x 20 Å x 20 Å unit cell with periodic boundary conditions. The structures are geometrically optimized until the residual force acting on each atom becomes less than 0.03 eV/Å.

## III. RESULTS

### III.A. MD simulations of sp-defect migration in an odd fullerene.

To obtain odd fullerenes, we have performed 70 MD simulation runs of 150 ns duration at temperature 2700 K with different amorphous clusters as an initial system. The detailed information about generation of amorphous carbon clusters is given in the Suppoting Information (see Figure S1). 9 clusters out of 70 transformed into odd fullerenes (with all $sp^2$ atoms except one sp atom) within this time: 1 $C_{59}$, 5 $C_{63}$, 2 $C_{65}$ and 1 $C_{69}$. About the same yield of even fullerenes, 8 out of 47 amorphous carbon clusters with an even number of atoms, was observed in our previous MD simulations within the simulation time of about 1500 ns at a lower temperature of 2500 K[19].

The detailed analysis of structural transformations observed on the atomic level in the previous MD simulations of transformation of amorphous carbon clusters into even fullerenes[19] allows us to conclude that the main atomistic mechanism at the last stage of formation of the $sp^2$ structure is insertion of atomic chains attached to the forming shell by both ends. Since the process of chain insertion is determined by the local structure of the forming fullerene shell it should be the same for shells which contain odd and even numbers of atoms. Thus, we do not repeat here the study of the $sp^2$-structure formation for odd fullerenes. Note only that no emission of atoms or molecules has been observed in the present MD simulations during formation of odd fullerene shells (identically to the previous result for even fullerenes[19]). Therefore, it can be expected that in typical experiments of fullerene synthesis, about half of the total number of the fullerenes will be odd and the other half even at the moment when formation of the $sp^2$ shell from the initially chaotic carbon system is complete.

Using reactive MD simulations, we have then studied the dynamics of bond rearrangements in odd fullerenes after the moment when the $sp^2$ structure of the fullerene shell is completed and only one sp atom remains in the shell. To circumvent difficulties related with the distinction between a vacancy and an adatom in the fullerene structure, a single sp atom is referred to here as an sp defect. In addition to the sp defect, the initial structure of the odd fullerenes contains one or several 7-rings, while the rest of the rings in the structure are 5- or 6-rings. The simulations are motivated by the idea that annealing of 7-rings is possible through autocatalysis of the SW reaction by a nearby sp atom, as proposed previously[4,42,43,46]. To check this hypothesis, we have performed MD runs of 600 ns duration at temperature 3000 K for all 9 odd fullerenes obtained. Different from formation of the fullerene shell from the amorphous carbon cluster, which starts from a fast decrease of the potential energy of the system, no decrease of the potential energy is observed here. In addition to the 7-rings existing in the initial structure, formation and annealing of 4- and 8-rings takes place through bond rearrangement reactions. Although 3-rings have been observed in reactive MD simulations at the initial stage of fullerene formation[71], 3-rings as well as 9-rings or other higher rings have not been revealed in our simulations. Since only 5- and 6-rings are present in a structure of abundant fullerene isomers we consider here the sum of the numbers of 4-, 7- and 8- rings, $N_d$, as the number of $sp^2$ structure defects different from the sp defect. This number fluctuates with time near the average value $N_d \sim 1 - 2$. The time dependences of the numbers of 4, 7- and 8- rings and potential energies of the fullerenes are shown in Figure S3 in Supporting information for all 9 considered odd fullerenes. Thus, we can conclude that several 4-, 7-, and 8- rings in addition to 5- and 6-rings are usually present in the odd fullerene structure at temperature 3000 K in thermodynamic equilibrium. To observe annealing of 4-, 7- and 8- rings, a lower temperature should be considered. Such studies, however, are beyond the possibilities of the present MD simulations due to the large computational time required. To obtain the statistics on the energetics and frequency of bond rearrangement reactions, 10 additional MD simulation runs of 600 ns duration at temperature 3000 K and 10 MD simulation runs of 300 ns duration at temperature 2400 K have been performed for the $C_{69}$ fullerene and 10 MD simulation runs of 300 ns duration at temperature 2400 K have been performed for the $C_{63}$ fullerene. Since the odd fullerenes considered (with several 4-, 7- and 8- rings in their structure) are in thermodynamic equilibrium at temperature 3000 K, we have averaged the simulation results over the runs at the same temperature and for the same number of atoms. We do not give below the information on the numbers of 5- and 6-rings since these numbers are close to that of the perfect fullerene and the numbers of different rings in the fullerene structure are related through the Euler formula. The calculated dependences of the numbers of 5-, 6-, and 7-rings are shown in Figure S4 in Supporting information for all 9 considered odd fullerenes.

One of the most interesting results of the simulations performed is the following. For all the considered odd fullerenes, we have observed numerous and diverse bond rearrangement reactions and they always involve the sp atom. That is no reactions occur in the major part of the fullerene shell which contains only $sp^2$ atoms. This is analogous to our previous MD study for even fullerenes (which contain only $sp^2$ atoms), where no bond rearrangements took place after the completion of shell formation at temperature of 2500 K for the total simulation time of about 10 μs[19]. The reactions observed here are divided into three types based on their relation with annealing of $sp^2$ defects (such as annealing of 7-rings or separation of 5-rings). The reactions of type I are those that change the number of atoms in rings of $sp^2$ structure and can lead to annealing or formation of $sp^2$ defects (such as the SW reaction facilitated by the nearby sp atom[37,38,41,42]). An example of such reaction is shown in Figure 1a. The reaction shown in Figure 1a is the first proposed autocatalytic reaction[42]. It is found among different reactions of type I in our MD simulations. The reaction of type II is migration of the sp defect and is shown in Figure 1b. This reaction is analogous to exchange mechanism of adatom migration on a surface and has been observed previously by the MD simulations only for a carbon nanotube with the end closed by a cap[72]. The reactions of type III involve short-living one-coordinated atoms or a chain of two sp atoms. The scheme of the chain formation is shown in Figure 1c. Note that such reactions are always followed by the reverse reaction within a short time period of less than 25 ps (that is within one step for structural analysis in our simulations). Here we focus on the results of the MD simulations and DFT calculations for sp-defect migration. The detailed schemes, statistics and activation barriers for the reactions of type I and III will be considered elsewhere.

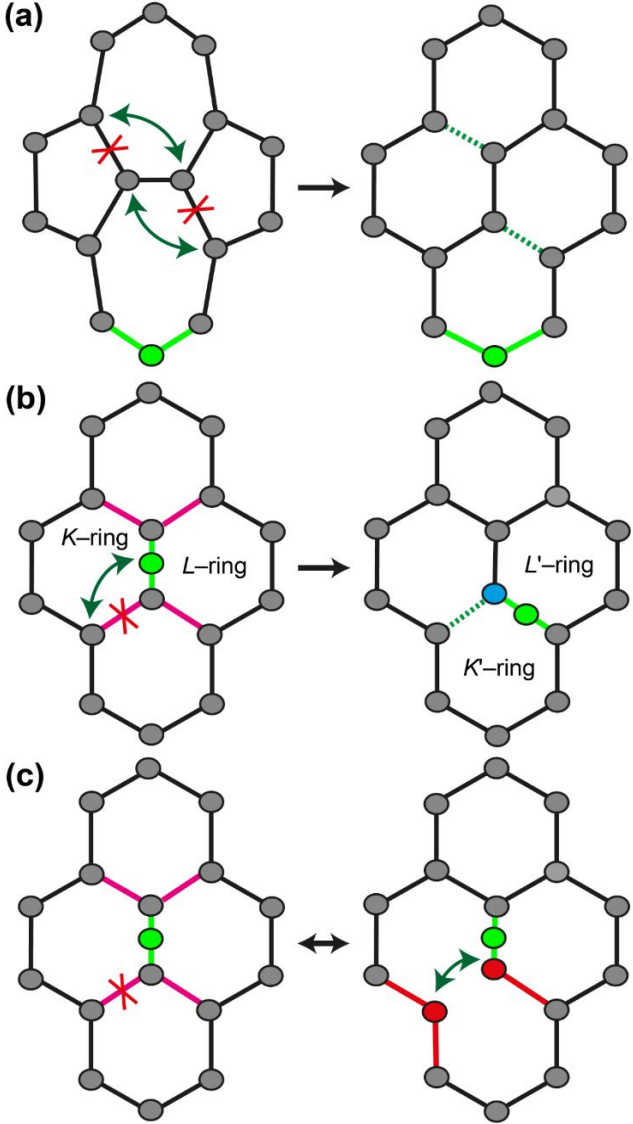

Figure 1. Schemes of reactions of type I (a), II (b) and III (c) observed in an odd fullerene. Forming and breaking bonds are indicated by dashed green lines and X, respectively. The green arrow points to atoms that form a new bond after breaking of the bond indicated by X. (a) and (b): the current sp atom is colored in green. (b) $K$, $L$ and $K'$, $L'$ are the numbers of $sp^2$ atoms in two rings which contain the bond replaced by the sp atom before and after the migration event, respectively (notation of migration events is described in the text). The former sp atom is colored in blue. For (b) and (c): 4 equivalent bonds that can be potentially broken in a single reaction are colored in magenta; (c) the atom which is sp defect before and after the reaction of type III is colored in green. Sp–atoms temporarily formed in the reaction of type III are colored in red.

The sp-defect migration event in an odd fullerene occurs as follows (see the scheme in Figure 1b). The sp atom has two nearest and four second-nearest $sp^2$ atoms. During the migration event, one of the four bonds between these nearest and second-nearest $sp^2$ atoms breaks (these bonds are coloured in magenta in Figure 1b). Simultaneously the new bond forms between the former sp atom and the second-nearest $sp^2$ atom with the broken bond. Thus, one of the nearest $sp^2$ atoms becomes the new sp defect. The additional sp atom in an even fullerene can be in two positions: above a bond between two atoms of $sp^2$ structure (that is two neighbouring atoms are 4-coordinated) and integrated into the fullerene shell instead of this bond (that is two neighbour atoms are $sp^2$ atoms)[42]. In our high-temperature MD simulations, the bond under sp atom is always absent. Following the previous papers devoted to DFT calculations of energetics of odd fullerenes[42,43,73], we describe the position of the sp atom using the notation $K$-$L$, where $K$, $L$ are the numbers of $sp^2$ atoms in two rings of $sp^2$ structure in the corresponding even fullerene which contains the bond replaced by the sp atom. For the sp-defect migration event, we introduce the notation $K$-$L \rightarrow K'$-$L'$, where the numbers $K$, $L$ and $K'$, $L'$ describe the structure before and after the sp-defect migration event, respectively (see Figure 1b). Evidently that the sp-defect migration event does not change the $sp^2$ structure of the corresponding even fullerene with the bond instead of the sp atom and thus cannot lead to annealing of $sp^2$ defects. Note that another possible bond rearrangement reaction which also preserves the $sp^2$-structure of the corresponding even fullerene, that is direct hopping of the same sp atom to the position instead of one of the neighbouring bonds (considered in DFT calculations[73]), has not been observed in our MD simulations.

The total number $N_m$ of detected sp-defect migration events and calculated average time $\Delta t_m$ between these events are presented in Table 1. Whereas more than 10000 migration events have been detected in the $C_{69}$ fullerene at temperature 3000 K during 6.6 μs, only 42 reactions of type I and 132 reactions of type III have been observed in the same 10 simulation runs. An example of the structure evolution of an odd fullerene during sp-defect migration and the number $N_2$ of former sp atoms from the beginning of the simulation run are presented in Figure 2. Figure 2 shows that about 1000 or 200 migration events are necessary for the sp defect to migrate through nearly the whole fullerene shell or half of the shell, respectively. The latter value can be used as an estimate of the average number of migration events required for the sp defect to arrive to a certain place of the fullerene shell after insertion of an additional atom into the structure of the even fullerene (for example, to meet an $sp^2$ defect which can be annealed with assistance of the sp atom).

All reactions of type I (which change the number of atoms in the rings of $sp^2$ structure) observed in our simulations occur with the help of the sp atom. Such reactions can lead in principle to annealing of 4-, 7- and 8-rings and separation of 5-rings, that is to annealing of $sp^2$ defects and, therefore, to selection of abundant isomers of fullerenes. Figure 3a shows that one ($N_d = 1$) or two ($N_d = 2$) 4-, 7- and 8-rings are present in most of the $C_{69}$ fullerenes at temperature 3000 K. Nevertheless, about a quarter of the fullerenes do not have these $sp^2$ defects (two rings which contain sp atom are not taken into account). According to the calculations[46], the presence of 7-rings in the $sp^2$ structure of an even fullerene leads to an increase of the fullerene potential energy. Since this increase is determined by the local atomic structure, we assume that the presence of 7-rings in the $sp^2$ structure of an odd fullerene should also increase its energy. To confirm this assumption, the potential energy at zero temperature has been calculated for all the structures of the $C_{69}$ fullerene observed in the MD simulations between subsequent bond rearrangement reactions. Figure 3b shows that the average potential energy of $C_{69}$ fullerene grows linearly upon increasing the number $N_d$ of 4-, 7- and 8-rings (about 0.7 eV per one ring). Thus, the odd fullerenes without 4-, 7- and 8-rings are more energetically favourable at zero temperature and therefore, the equilibrium portion of odd fullerenes without such rings should increase upon decreasing temperature. The energy decrease upon $sp^2$-defect annealing in an odd fullerene has been show previously only by the example of 5-rings separation in $C_{61}$ fullerene[4,42,43].

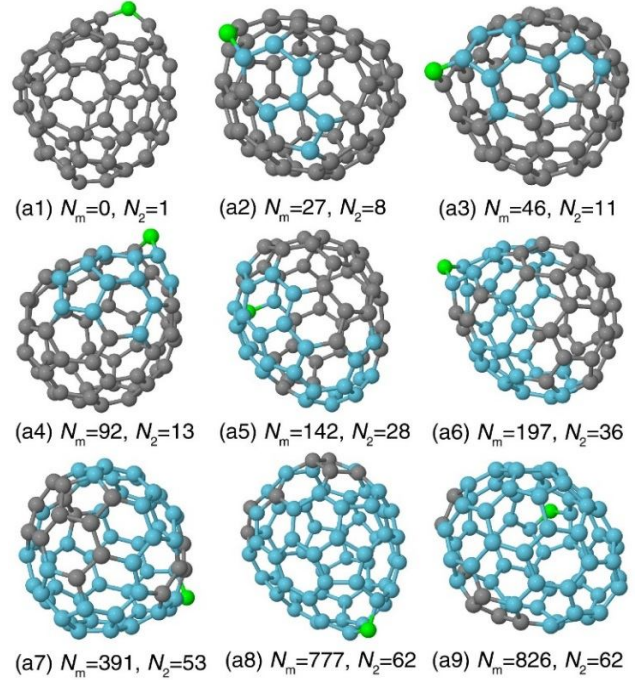

Figure 2. Simulated structure evolution showing the sp-defect migration in the $C_{69}$ fullerene at temperature 3000 K: (a1) 0 ns (initial structure), (a2) 17.05 ns, (a3) 28.4 ns, (a4) 62.35 ns, (a5) 113.6 ns, (a6) 156.95 ns, (a7) 285.2 ns, (a8) 536.9 ns and (a9) 600 ns. The current and former sp atoms are coloured in green and blue, respectively. The number of former sp atoms, $N_2$, and total number of migration events, $N_m$, are indicated.

Table 1. The total number $N_m$ of detected sp-defect migration events and calculated average time $\Delta t_m$ between these events for odd fullerenes consisting of $N_a$ atoms at temperature $T$ and during simulation time $t$.

| $N_a$ | $t$, μs | $N_m$ | $T$ (K) | $\Delta t_m$ (ns) |
|---|---|---|---|---|
| 59 | 0.6 | 358 | 3000 | 1.0±0.2 |
| 63 | 3.0 | 4618 | 3000 | 0.62±0.07 |
| 65 | 1.2 | 1883 | 3000 | 0.5±0.08 |
| 69 | 6.6 | 10061 | 3000 | 0.65±0.03 |
| 63 | 3.0 | 257 | 2400 | 6.9±0.8 |
| 69 | 3.0 | 63 | 2400 | 27±6 |

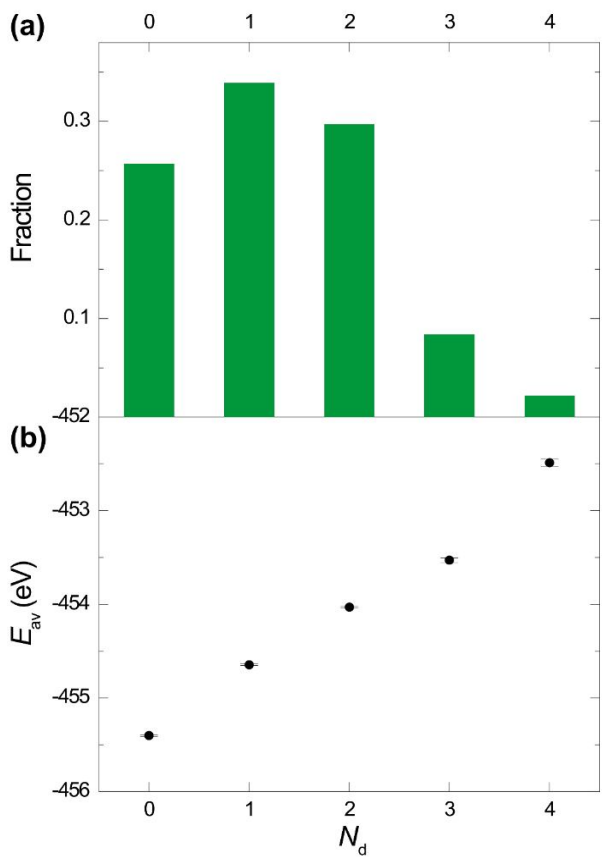

Figure 3. (a) Calculated fraction of $C_{69}$ fullerenes with a certain number $N_d$ (which is the sum of the numbers of 4-, 7- and 8-membered rings) at temperature 3000 K (rings which contain sp atom are not taken into account) and (b) average potential energies $E_{av}$ of these fullerenes at temperature 0 K. Both quantities are averaged over all structures observed between subsequent bond rearrangement reactions.

Two sp defects can coexist in an even fullerene simultaneously just after the fullerene shell formation or upon insertion of two single carbon atoms from the surrounding vapour to different places of the shell. In this case, annihilation of two sp defects is possible once they meet during the migration. Such a case of annihilation of an sp-defect pair has been found in the additional analysis of old MD simulations runs for high-temperature transformation of even amorphous carbon clusters into fullerenes performed in our previous paper[19]. The structural transformation and reaction scheme for the $C_{66}$ fullerene are presented in Figure 4.

It is of interest that just after annihilation of the sp defects, in this particular case, the fullerene has the perfect $sp^2$ structure which consists only of 5- and 6-rings. Previously the reaction of annihilation of two sp defects was observed in a closed cap of a carbon nanotube in MD simulations of carbon nanotube growth[72]. Note that reactions of annihilation of a sp-defect pair presented here and in[72] have different configurations of formed and broken bonds. The annihilation of a sp-defect pair is the third type of final reactions that lead to formation of the fullerene shell which consists only of $sp^2$ atoms in addition to the previously reported detachment of carbon chains attached to the shell by one end[25] and insertion of carbon chains attached by both ends into the fullerene shell structure[19].

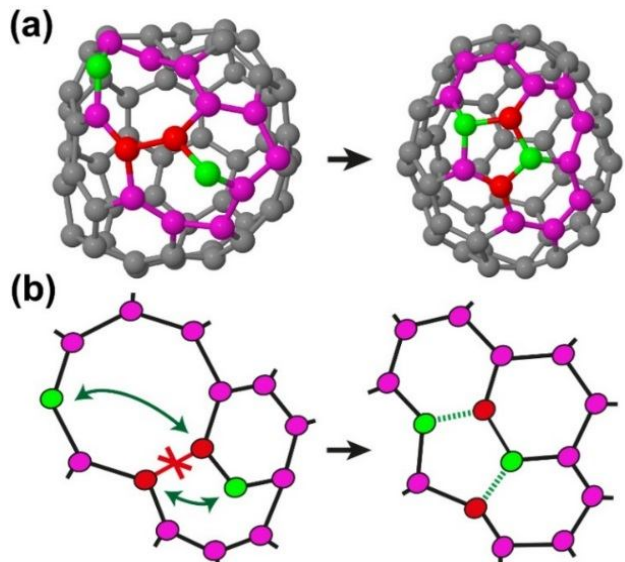

Figure 4. Calculated structures (a) and corresponding scheme (b) showing the reaction of annihilation of a nsp-defect pair in the $C_{66}$ fullerene. The sp atoms are colored in green. The atoms with the breaking bond are colored in red. The other atoms in the rings that include atoms with forming and breaking bonds and shown in the scheme are colored in magenta. The forming bonds are indicated by dashed green lines. The breaking bond is indicated by X. The green arrows point to the atoms that form new bonds.

The simulations performed in the present paper do not reveal emission of the extra sp atom or any carbon molecule from the odd fullerene. This agrees with the estimates based on the experimental studies of $C_{60}$-$I_h$ with $C^+$ collisions that the $C^+_{61}$ fullerene is stable with respect to emission of the extra atom at the millisecond scale at temperature about 1500 K[49]. Therefore, the barrier of emission of the extra atom is higher than the barriers of the reactions of bond rearrangements in odd fullerenes observed in our MD simulations, while attachment of a single carbon atom to the $sp^2$ structure of an even fullerene is barrierless[46]. This conclusion is in accordance with the DFT calculations suggesting that about 3 eV[42] and 4.8 eV[46] are necessary for removal of the sp atom from the $C_{61}$ and $C_{69}$ fullerenes, respectively. Attachment of single carbon atoms and annihilation of sp-defects pairs have been also included into the set of reactions proposed to explain formation of the abundant isomer $C_{70}$-$D_{5h}$ observed in laser ablation of the $C_{60}$-$I_h$ fullerene simultaneously with amorphous $^{13}C_4$. We believe that attachment of single carbon atoms, sp-defect migration to the defected place of the $sp^2$-structure, $sp^2$ defect annealing assisted by the sp atom (such as autocatalysis of the SW reaction) and subsequent annihilation of pairs of sp-defects should considerably contribute to the atomistic mechanism of formation of abundant isomers of fullerenes under diverse conditions of fullerene synthesis.

**III.B. DFT study of sp-defect migration energetics.**
The MD simulations of kinetic characteristics of the sp-defect migration are supplemented by the DFT study of energetics of migration events. 10 migration events in the $C_{69}$ fullerene with different positions of the sp atom observed in the MD simulations at 3000 K have been randomly chosen (see Figure 1 and the related text for the notation of the migration event). The values of the barriers $E_a$ and energy changes $\Delta E = E_2 - E_1$, where $E_1$ and $E_2$ are energies of an odd fullerene before and after the migration event, respectively, have been obtained by the DFT calculations and using the empirical potential (see Table 2). Analogously to previous DFT studies of energetics of odd fullerenes, we describe the position of the sp atom as above or instead the bond between two nearest sp² atoms[42,43,73]. The distances $d_1$ and $d_2$ between these sp² atoms before and after the migration event, respectively, obtained by the DFT calculations are given in Table 2. The structures of the fullerene before and after the considered migration events are presented in Figure S3 of Supplementary Information.

Two cases have been found for the position of the sp atom relative to the sp² structure of the corresponding even $C_{68}$ fullerene. For the majority of the considered isomers of the $C_{69}$ fullerene, the sp atom replaces the bond between two nearest sp² atoms. However, isomers with the position of the sp atom above the bond with the length about 1.5 – 1.6 Å have been also obtained. This is consistent with the result of the previous DFT studies for the $C_{61}$ fullerene with the sp atom in the 5-6 and 6-6 positions relative to the sp² structure of the $C_{60}$-$I_h$ fullerene[42,73]. An isomer with the sp atom above the bond has been found for 5-6 position of the sp atom[42,73] and isomers with the sp atom instead of the bond have been observed both for the 5-6[73] and 6-6[42,73] positions.

The comparison of energy changes $\Delta E$ for migration events calculated by different methods shows excellent agreement between the results of the DFT calculations and those obtained with the empirical potential. Namely, the average absolute value of the differences between the energy changes of 0.153 eV computed with the VASP code (used to fit the parameters of the potential[56]) and the potential is even less than the average difference of 0.165 eV for the DFT approaches using the VASP and Priroda codes. However, the potential tends to overestimate the activation barriers of the sp-defect migration events by 1.2 eV on average. Such an overestimation of the barriers by the potential used means that the sp-defect migration should occur even easier (with a smaller time interval between subsequent migration events) than in our MD simulations. The absolute value of the energy change averaged over 10 considered migration events is $\Delta E$ = 0.4 ± 0.1, 0.45 ± 0.05 and 0.5 ± 0.1 eV for the calculations with the empirical potential, Priroda code and VASP code, respectively. The obtained values of $\Delta E$ are in good agreement with the previous DFT calculations giving 0.26 eV[42,43] and 0.3 eV[73] for the energy difference between isomers of $C_{61}$ fullerene with an sp atom in the 5-6 and 6-6 positions relative to the sp² structure of the $C_{60}$-$I_h$ fullerene. Our DFT calculations predict that the barriers of the sp-defect migration events should range from 1.19 to 2.19 eV with the average value of 1.6 ± 0.1 eV (see Table 2). It is interesting that the smallest DFT values of 1.19 – 1.44 eV correspond to the barriers of hopping of the same sp atom in the odd $C_{61}$ fullerene between the 5-6 and 6-6 positions[73]. Such a hopping, which is another possible way to switch the position of the sp atom relative to the $sp^2$ structure of the corresponding even fullerene (in addition to the sp-defect migration), has not been observed in our MD simulations. However, a competition between the sp-defect migration and sp-atom hopping can take place depending on the size and structure of the odd fullerene and temperature. Note that the computed barriers of sp-atom hopping and sp-defect migration are comparable to the DFT results of 1.0 – 1.3 eV for vacancy migration in graphene (see Ref. 74 for a review). However, they are considerably greater than the barriers 0.35 – 0.53 eV reported for adatom migration in graphene[74].

**Table 2.** Energy characteristics for the sp-defect migration events in the $C_{69}$ fullerene, barriers $E_a$ (in eV), energy changes $\Delta E$ (in eV), and distances $d_1$ and $d_2$ (in Å) between the two sp² atoms which are the nearest neighbours of the sp atom before and after the migration event, respectively, obtained using the interatomic potential and the DFT calculations with the Priroda and VASP codes. The notation $K$-$L \rightarrow K'$-$L'$ of the migration event is described in Figure 1 and the related text. The structures of the fullerene before and after the migration event shown in Figure S2 of Supplementary Information are indicated.

| Figure S2 | $K$-$L \rightarrow K'$-$L'$ | Potential | | DFT Priroda | | | | DFT VASP | | |
|---|---|---|---|---|---|---|---|---|---|---|
| | | $E_a$ | $\Delta E$ | $E_a$ | $\Delta E$ | $d_1$ | $d_2$ | $\Delta E$ | $d_1$ | $d_2$ |
| a | 5-6 → 5-6 | 3.3 | 0.9 | 2.19 | 0.61 | 2.22 | 1.52 | 0.95 | 2.24 | 1.59 |
| b | 5-5 → 6-6 | 2.85 | 0.38 | 1.38 | 0.52 | 2.23 | 2.33 | 0.45 | 2.24 | 2.1 |
| c | 5-6 → 6-6 | 3.08 | 0.41 | 1.72 | 0.26 | 2.18 | 2.21 | 0.24 | 2.18 | 2.23 |
| d | 6-6 → 5-6 | 2.58 | 0.35 | 1.73 | 0.59 | 2.16 | 2.00 | 0.67 | 2.21 | 1.90 |
| e | 6-6 → 6-6 | 2.44 | 0.16 | 1.40 | 0.41 | 2.18 | 2.15 | 0.54 | 2.24 | 2.13 |
| f | 6-5 → 6-5 | 3.52 | 0.16 | 1.71 | 0.48 | 2.39 | 2.38 | 0.15 | 2.24 | 2.43 |
| g | 5-6 → 5-5 | 2.23 | -0.03 | 1.33 | -0.15 | 2.05 | 2.21 | -0.07 | 2.23 | 2.23 |
| h | 5-6 → 6-6 | 3.51 | -0.19 | 1.19 | -0.25 | 2.38 | 2.42 | -0.43 | 2.23 | 2.44 |
| i | 5-6 → 5-6 | 2.54 | -0.52 | 1.86 | -0.47 | 1.63 | 2.12 | -0.42 | 1.79 | 2.12 |
| j | 6-6 → 6-6 | 2.32 | -1.01 | 1.67 | -0.79 | 1.52 | 2.22 | -1.16 | 1.59 | 2.24 |

Let us now estimate the time between subsequent migration events during the fullerene formation under experimental conditions based on the results of the MD simulations and DFT calculations. Using the average barrier of the migration event $E_a$ = 2.8 eV obtained for the $C_{69}$ fullerene using the empirical potential and the average time between subsequent migration events $\langle \tau \rangle$ = 0.65 ns at 3000 K obtained in the MD simulations (see Table 1) in the Arrhenius equation $\langle \tau \rangle = \tau_0 \times \exp(E_a / k_B T)$, where $k_B$ is the Boltzmann constant, we get a rough estimate of the pre-exponential factor $\tau_0 \sim 10^{-14}$ s. Then from the same Arrhenius equation with the estimated pre-exponential factor and the average barrier of the migration event $E_a$ = 1.6 eV following from our DFT calculations, we find that the time intervals between subsequent migration events at temperatures 1000 K and 1500 K should be 1 μs and 10 ns, respectively. The time of formation of the $C_{60}$ fullerene in laser ablation of graphite in a furnace filled with a buffer gas exceeds 0.4 ms[75]. Thus, more than 40000 and 400 migration events occur in this process at temperatures 1500 and 1000 K, respectively. According to our MD simulations, several hundreds of migration events are necessary for the sp defect to migrate through about a half of the fullerene shell (see Figure 2), which on average corresponds to reaching another defect (such as 7-ring and so on) that can be annealed by autocatalysis (i.e. any reaction of bond rearrangements promoted by the sp atom). Therefore, the estimated number of migration events under experimental conditions is sufficient to anneal defects of $sp^2$ structure during the fullerene formation and to provide that the most energetically stable isomer becomes the most abundant one.

IV. CONCLUSIONS

The dynamics of bond rearrangements in odd fullerenes $C_{59}$, $C_{63}$, $C_{65}$ and $C_{69}$ at temperatures 2400 – 3000 K has been studied by reactive MD simulations. For the most part of the time, the structure of the studied odd fullerenes contains a single sp atom among $sp^2$ atoms. Since the distinction between adatom or vacancy is not evident for the general case of the odd fullerene structure, the sp atom is referred here as sp defect. About 20000 bond rearrangement reactions nearby the sp defect have been observed, whereas no bond rearrangements have been detected for the rest of the $sp^2$ structure of the odd fullerene. The reactions revealed have been divided into three types based on their relation with annealing of defects of $sp^2$ structure: reactions of type I are those that change the number of atoms in rings of $sp^2$ structure, type II is sp-defect migration, and reactions of type III involve short-living one-coordinated atoms or two additional sp atoms and are always followed by the fast reverse reaction.

Let us discuss the results of the MD simulations in relation with the atomistic mechanism of formation of abundant isomers of fullerenes. As discussed in Introduction, the most realistic hypothesis for such a mechanism is bond rearrangement reactions assisted by an extra sp carbon atom via autocatalysis[4,42,43,46,48] (as confirmed by recent experiment[4]). These reactions (referred here as type I) take place in odd fullerenes and anneal defects of $sp^2$ structure present in the initial fullerene shell formed through self-organization. 42 reactions of type I have been detected in the MD simulations performed. At the same time, the MD simulations show that there are no bond rearrangement reactions in the major part of the odd fullerene shell consisting only of $sp^2$ atoms. Thus, we confirm the hypothesis that annealing of defects of $sp^2$ structure should proceed through bond rearrangement reactions assisted by the sp atom.

While the sp atom can play the main role in annealing of defects of $sp^2$ structure, little attention has been given previously to the ways of how the sp atom approaches such defects and is eliminated after annealing of these defects. The MD simulations performed allow us to propose that the sp-defect migration is the way to deliver the sp atom close to the defects of fullerene $sp^2$ structure. The stochastic nature of the sp-defect migration is clear from the fact that almost all atoms of the odd fullerene become sp defect at least once during the simulation run. Let us discuss possible routes of the sp atom removal from the odd fullerene. In our MD simulations, we have detected in total about 20000 sp-defect migration events for all the considered odd fullerenes and simulation runs. At the same time, no emission of the sp atom has been observed. Subsequent attachment of single carbon atoms and annihilation of a pair of sp defects have been proposed to explain formation of the abundant isomer $C_{70}$-$D_{5h}$ during laser ablation of the $C_{60}$-$I_h$ fullerene simultaneously with amorphous $^{13}C$[4]. The annihilation of the pair of sp defects has been found here by the additional analysis of the old MD simulation results on high-temperature transformation of even amorphous carbon clusters into fullerenes[19]. The sp-defect migration can also play an important role for structural transformation of fullerenes in transmission electron microscope. Through migration and annihilation of sp-defects, self-healing of the fullerene can take place after knock-on removal of atoms. This means that the sp² structure of the fullerene can be preserved under irradiation with the fullerene size gradually decreasing down to the minimal possible one.

Previously the energy decrease in annealing of defects of sp² structure in odd fullerenes due to reactions assisted by the sp atom (reactions of type I) has been shown only by the example of separation of 5-rings in the $C_{61}$ fullerene [4,42,43]. Here potential energies at zero temperature have been calculated for more than 11000 isomers of the $C_{69}$ fullerene observed in the MD simulations. It is found that on average the potential energy of this fullerene is reduced upon decreasing the total number of 4-, 7- and 8-rings. This means that annealing of 4-, 7- and 8-rings through reactions of type I in odd fullerenes is accompanied by an energy decrease. Note that the differences between the values of the energy change in migration events obtained using the REBO-1990EVC potential and by the DFT calculations lie within 0.15 eV. This demonstrates that the potential adequately describes the correlation between the potential energy of odd fullerenes and the presence of 4-, 7- and 8-rings in their structure.

The barrier of sp-defect migration has been investigated through the DFT calculations for 10 different migration events and the average value of 1.6 eV has been obtained. This value of the barrier and the average time between subsequent migration events have been used to estimate the average time between subsequent migration events under experimental conditions at temperatures 1000 – 1500 K using the Arrhenius equation. The estimates show that the rate of sp-defect migration is more than sufficient to deliver the sp atom close to sp²-structure defects within the fullerene formation time and multiple reactions of annealing of sp²-structure defects assisted by the sp atom are possible.

Thus, based on the MD simulations and DFT calculations performed, we suggest to supplement the paradigm of formation of the initial fullerene shell with sp²-structure defects via self-organization by the four-stage atomistic mechanism of annealing of sp²-structure defects. Namely, we propose that 1) attachment of single carbon atoms, 2) sp-defect migration to sp²-structure defects, 3) sp² defect annealing assisted by the sp atom (such as autocatalysis of the SW reaction) and 4) subsequent annihilation of the pair of sp defects can considerably contribute to the atomistic mechanism of formation of abundant isomers of fullerenes under diverse conditions of fullerene synthesis where the abundant isomers are observed.

## ASSOCIATED CONTENT

**Supporting Information:** Scheme of generation of amorphous carbon clusters from which odd fullerenes were obtained, figures illustrating correlation between the total number of 4-, 7-, and 8-rings and the potential energy of odd fullerenes, changes in the number of 5-, 6- and 7-rings with time and examples of structural changes in different sp-defect migration events in the $C_{69}$ fullerene.

## AUTHOR INFORMATION

**Corresponding Author**

* E-mail: popov-isan@mail.ru.

## ACKNOWLEDGMENT

ASS, YGP, AMP and AAK acknowledge the Russian Foundation of Basic Research (Grants 18-02-00985 and 18-52-00002). This work has been carried out using computing resources of the federal collective usage center Complex for Simulation and Data Processing for Mega-science Facilities at NRC "Kurchatov Institute", http://ckp.nrcki.ru/.

# Supporting Information for

# Molecular Dynamics Study of sp-Defect Migration in an Odd Fullerene: Possible Role in Synthesis of Fullerenes

Alexander S. Sinitsa[a], Irina V. Lebedeva[b], Yulia G. Polynskaya[d],

Andrey M. Popov[c,1], and Andrey A. Knizhnik[d]

[a] National Research Centre "Kurchatov Institute", Kurchatov Square 1, Moscow 123182, Russia.

[b] CIC nanoGUNE, Avenida de Tolosa 76, San Sebastian 20018, Spain.

[c] Institute for Spectroscopy of Russian Academy of Sciences, Fizicheskaya Street 5, Troitsk, Moscow 108840, Russia.

[d] Kintech Lab Ltd., 3rd Khoroshevskaya Street 12, Moscow 123298, Russia.

**Contents**

[1]Corresponding author. Tel. +7-909-967-2886 E-mail: popov-isan@mail.ru (Andrey Popov)

# Generation of the initial structure of amorphous carbon clusters

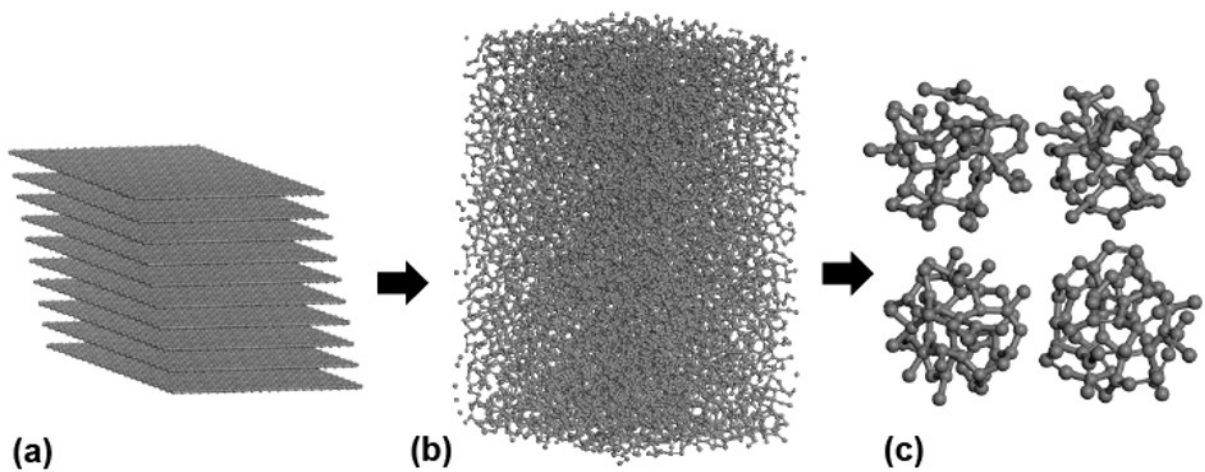

**Figure S1**. Scheme of generation of the amorphous carbon structure and initial amorphous carbon clusters: (a) 10 graphite layers of 800 atoms before annealing; (b) amorphous carbon after annealing at temperature 8000 K and quenching to room temperature; (c) examples of amorphous carbon clusters cut from the amorphous structure obtained.

The common method for preparation of amorphous carbon structures in molecular dynamics (MD) simulations is annealing of several graphite layers or diamond structure at high (6000-7000 K) temperature followed by quenching to room temperature[1-5]. We performed MD annealing of ten graphite layers (8000 carbon atoms in 200x200x200 Å supercell) structure at 8000 K, followed by rapid (10 K/fs) quenching to room temperature. The amorphous carbon cluster was cut from the amorphous structure obtained using the following algorithm: firstly, the "center" of the cluster was chosen randomly in the quenched amorphous carbon bulk, then atoms farther than a predetermined distance from this chosen "center" (4.5 Å for clusters in our simulations) were removed to obtain a ball of amorphous carbon. If the total number of atoms was even, one atom was deleted, so all clusters contain only odd number of atoms. Typically, the cluster obtained contained from 57 to 73 atoms.

To get odd fullerenes, we have performed 70 MD simulation runs of 150 ns duration at temperature 2700 K with different amorphous clusters as an initial system in the same 200x200x200 Å supercell. 9 clusters out of 70 completely transformed into odd fullerenes (with all $sp^2$ atoms except one sp atom) within this time: 1 $C_{59}$, 5 $C_{63}$, 2 $C_{65}$ and 1 $C_{69}$. As for the rest of the simulation runs, the same transformation process was observed, although one or a few chains remained attached to the hollow shell with the $sp^2$ structure after 150 ns of the simulation time. About the same yield of even fullerenes, 8 out of 47 amorphous carbon clusters with an even number of atoms, was observed in our previous MD simulations within the simulation time of about 1500 ns at a lower temperature of 2500 K[6].

# Correlation between the total number of 4-, 7-, and 8-rings and the potential energy of odd fullerenes

No evident decrease of the total number $N_d$ of 4-, 7-, and 8-rings (rings which contain the sp atom are not taken into account) with time has been observed during a single simulation run at temperature 3000 K. In the MD simulations performed, the average number of such rings is about 1-2 in each structure. The calculated dependences of the potential energy show a clear correlation between the potential energy and the number of defects. In general, odd fullerenes with fewer numbers of 4-, 7-, and 8-rings are more energetically favorable and vice versa (see, for example, Figure S2d,a,g,i).

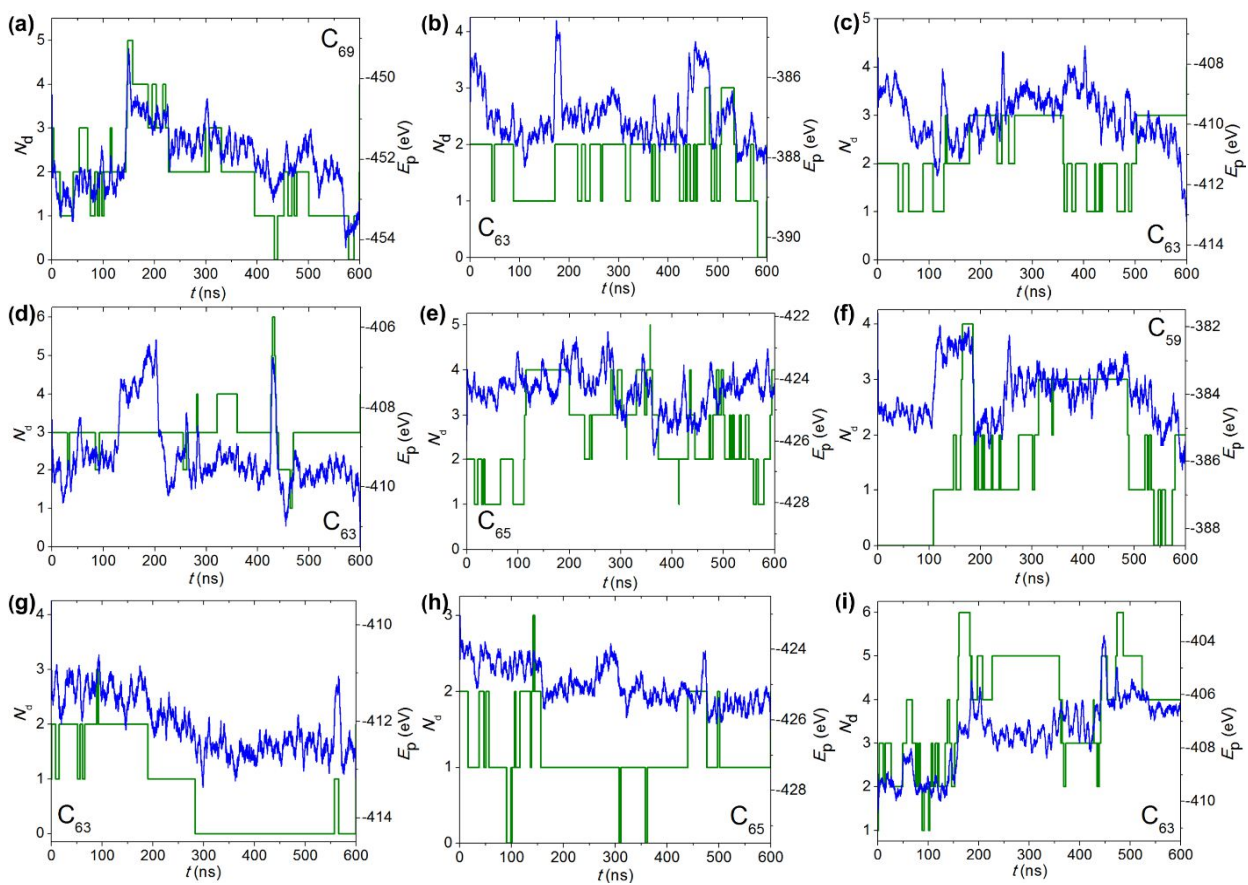

**Figure S2.** (a)-(i) Calculated dependences of the total number $N_d$ of 4-, 7-, and 8-membered rings (two rings which contain the sp atom are not taken into account), (left axis, green line) and the potential energy of odd fullerenes, $E_p$, (right axis, blue line) on time $t$ for 9 odd fullerenes considered. The considered odd fullerenes are indicated. The calculated values are averaged for each time interval of 5 ns. The correspondence between the fullerene and the panel is the same as in Figure S4.

**Structural changes in single sp-defect migration events**

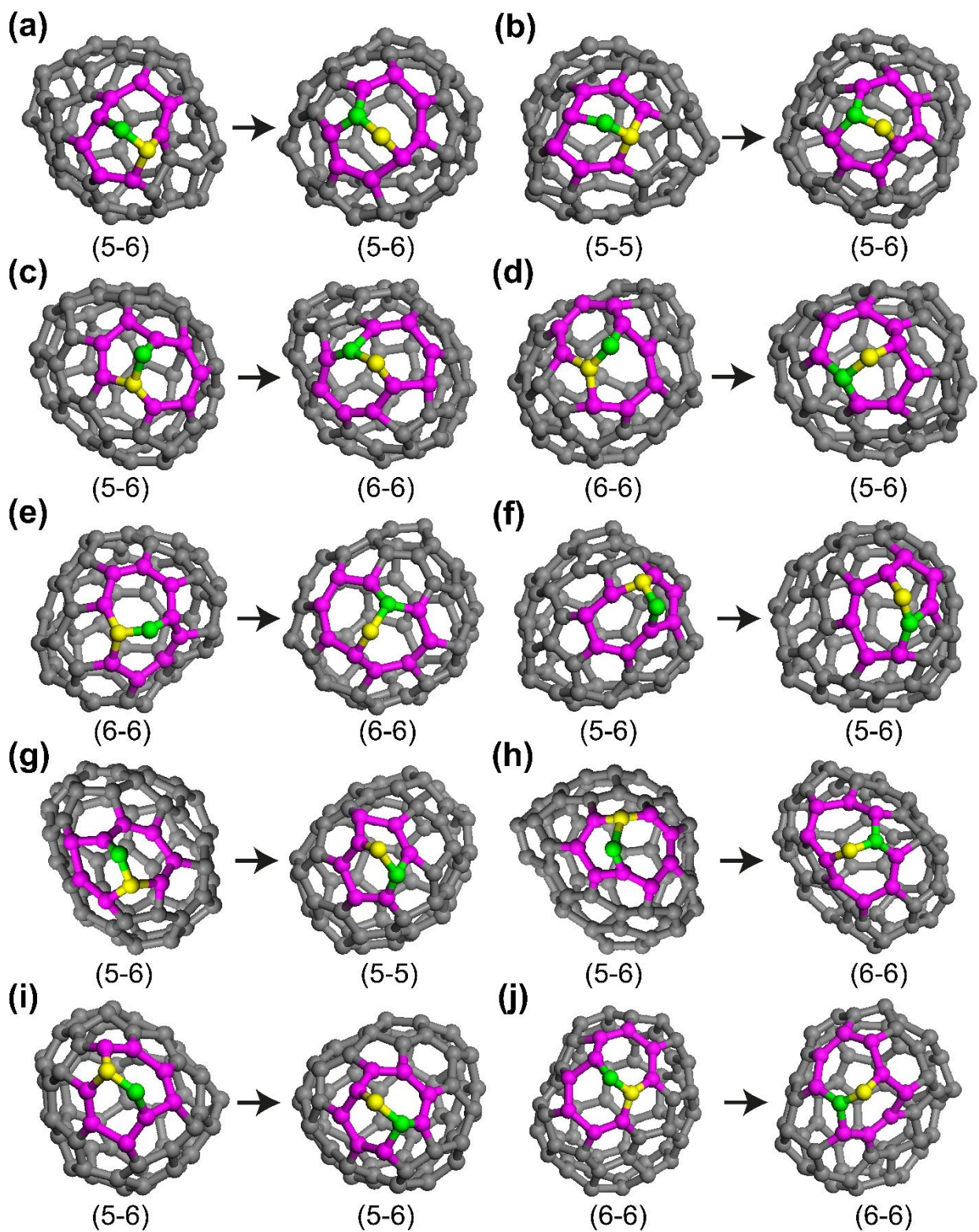

**Figure S3.** Structural changes observed indifferent sp defect migration events in the $C_{69}$ fullerene (a)-(j). Atom which have the sp hybridization before and after the migration event are colored in green and yellow, respectively, other atoms in the rings involved in the migration event are colored in magenta. The notation for the sp atom position is indicated (the notation is described in the text of the article). Calculated energy changes and barriers corresponding to the migration events shown in panels (a)-(j) are given in Table 2.

# Changes in the number of 5-, 6-, and 7-rings with time

Calculated dependences of the numbers of 5-rings, 6-rings and 7-rings in the odd fullerene structure on time are shown in Figure S4. The majority of the changes of these numbers are related with the temporarily appearance of the sp atom in the ring structure of a local region of an odd fullerene due to sp-defect migration event (reaction of type II). The $sp^2$ structure of this local region is the same before and after appearance of the sp atom in this region. Changes related with reactions of types I and III occur rarely.

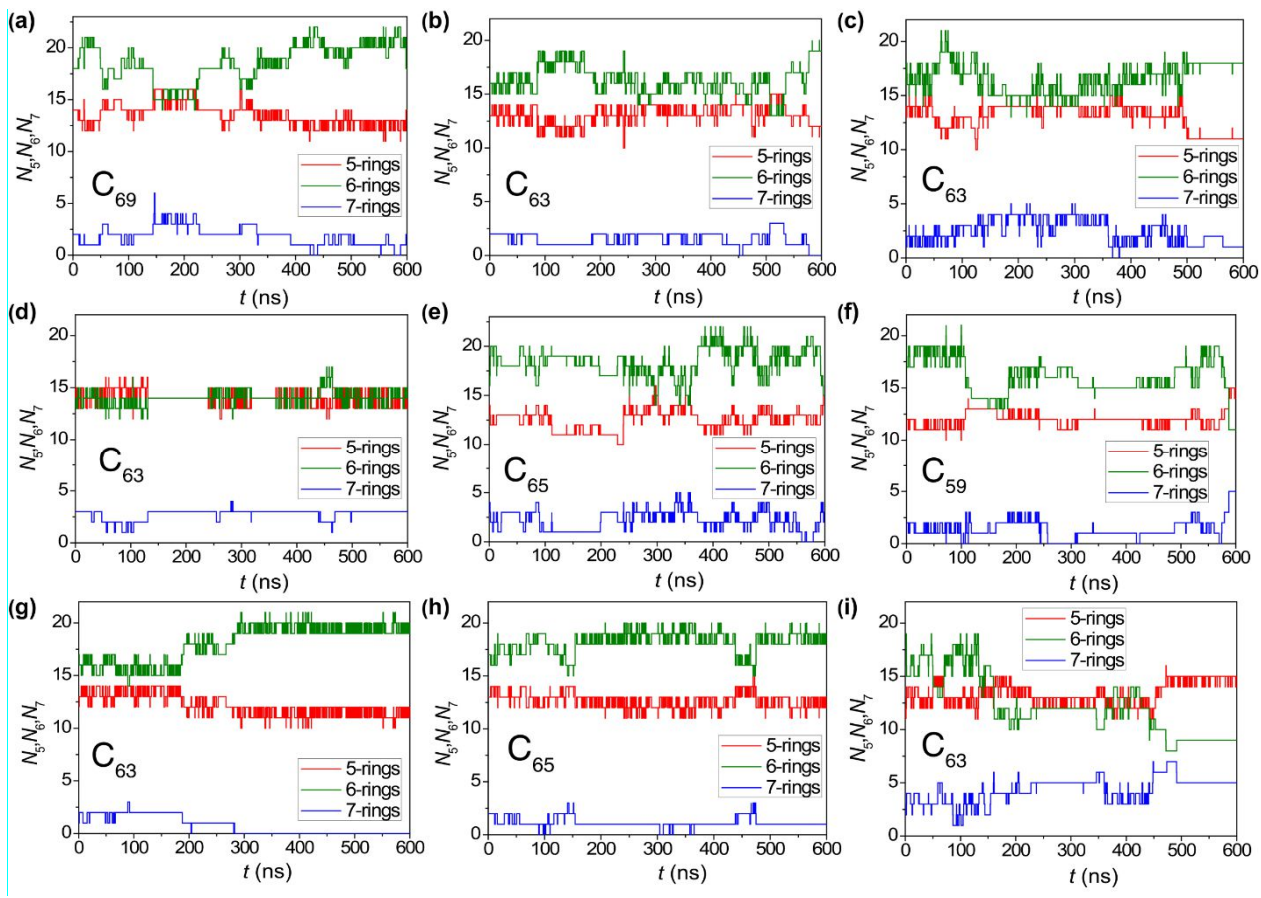

**Figure S4.** (a)-(i) Calculated dependences of the number $N_5$, $N_6$ and $N_7$ of 5-, 6-, and 7-membered rings (red, green and blue line, respectively) on time $t$ for 9 odd fullerenes considered. The considered odd fullerenes are indicated. The correspondence between the fullerene and the panel is the same as in Figure S2.

**Description of the video files attached**

Attached video: an example of simulated structure evolution showing the sp-defect migration in the $C_{69}$ fullerene at temperature 3000 K during 150 ns. The current and the former sp atoms are colored in green and blue, respectively. Temporarily emerging and disappearing other sp atoms are colored red. The corresponding simulation time in picoseconds and the number of the former sp atoms which acted like the migrating sp defect, $N_2$, is indicated below.

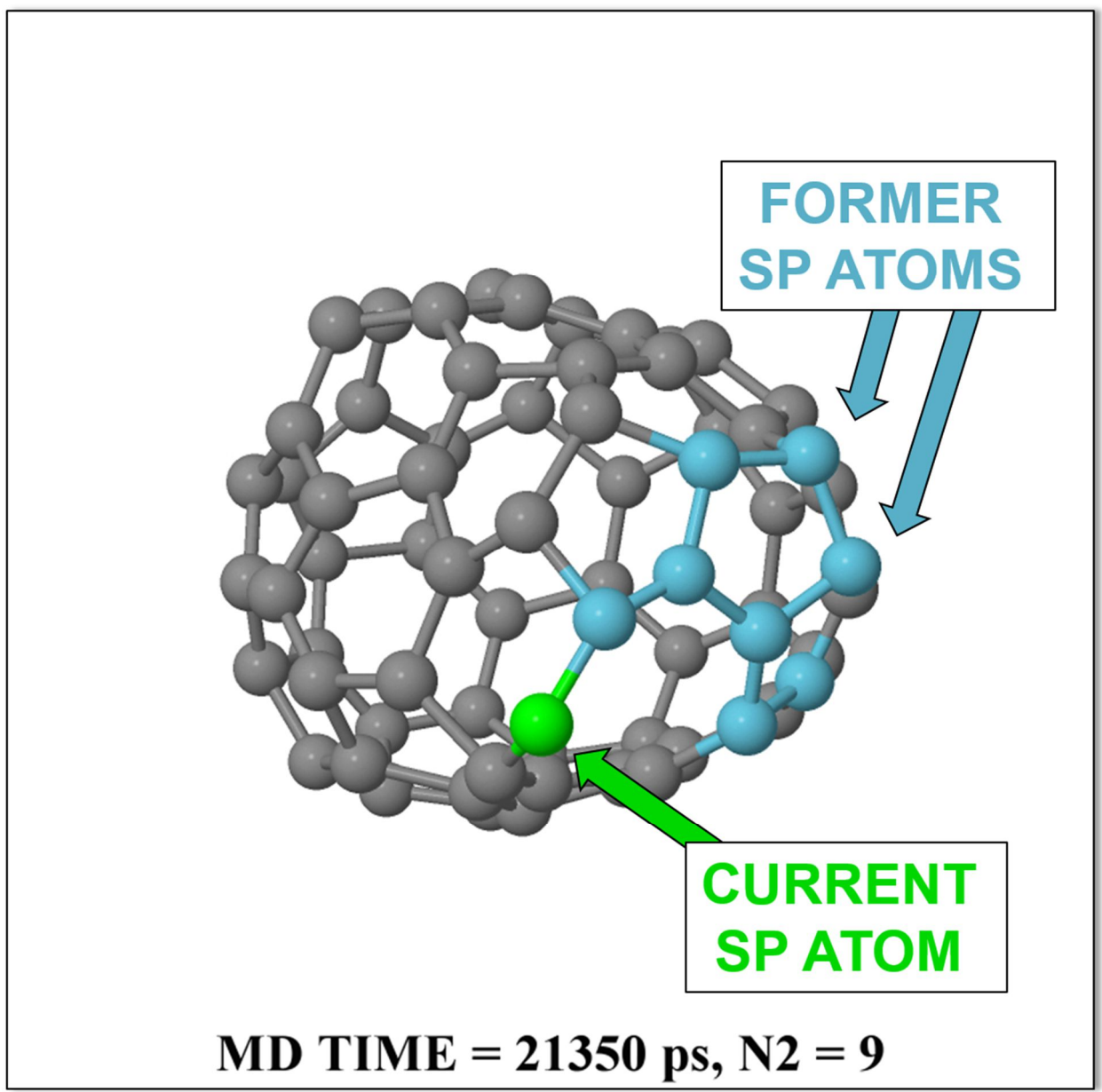